\documentclass[twoside]{article}
\usepackage{fleqn,espcrc2}

\newcommand{\AmS}{{\protect\the\textfont2
  A\kern-.1667em\lower.5ex\hbox{M}\kern-.125emS}}
\input{tcilatex}

\begin{document}

\title{Photoinduced absorption from localized intra-gap states.}
\author{V.V. Kabanov$^{\mathrm{a}}$, J. Demsar$^{\mathrm{a}}$ and D. Mihailovic%
\address{Josef Stefan
Institute 1001, Ljubljana, Slovenia.}}

\begin{abstract}
A model is developed for photoinduced absorption from localized states
observed in femtosecond pump-probe experiments in high-T$_{c}$
superconductors and other materials. The dynamics of localized carriers are
described in terms of phenomenological approach similar to that originaly
proposed by Rothwarf and Taylor. Expanding the relaxation rate in powers of
the order parameter we have shown that density of localized carriers is
sensitive to T$_{c}$. From the analysis of the experimental data on YBa$_{2}$%
Cu$_{3}$O$_{7-x}$ and K$_{0.3}$MoO$_{3}$ we conclude that significant
intra-gap density of localized states exists in these materials. Temperature
dependence of the density of photoexcited localized carriers in underdoped
YBa$_{2}$Cu$_{3}$O$_{7-x}$ and in K$_{0.3}$MoO$_{3}$ is consistent with the
observation of the pseudogap above T$_{c}$

\vspace{1pc}
\end{abstract}

\maketitle


\section{Introduction}

Recent pump-probe experiments on cuprate superconductors \cite
{Han,Stevens,Demsar,kdbm} and some other materials \cite{bb} have shown that
a photoinduced change in absorption or reflection can be observed at low
temperatures and especially for $T<T_{c}$. The effects are believed to be
caused by excited state absorption of the probe pulse from photoexcited
quasiparticle (QP) states \cite{kdbm} and theoretical analysis of the
response was found to be in good agreement with experimental data on YBa$%
_{2} $Cu$_{3}$O$_{7-\delta }$ over a wide range of doping \cite{kdbm}.
However, in addition to the QP response which occurs on the picosecond and
subpicosecond timescale, a distinct slower response was also consistently
observed in YBa$_{2}$Cu$_{3}$O $_{7-\delta }$ (YBCO) \cite{Stevens}, Bi$_{2}$%
Sr$_{2}$CaCu$_{2}$O$_{8}$ (BISCO) and Bi$_{2}$Y$_{x}$Ca$_{1-x}$SrCu$_{2}$O$%
_{8}$ \cite{Thomas} and more recently in the charge-density-wave (CDW) quasi
one-dimensional Peierls insulator K$_{0.3}$MoO$_{3}$\cite{bb}. It was shown
to be of non-thermal origin (detailed analysis is given in Ref. \cite{Thomas}%
) and occurs on a timescale of 10$^{-8}$ s or longer. Its anomalous T
-dependence in the two systems \cite{Stevens,bb} lead the authors to the
suggestion that the signal is due to localized states near the Fermi energy.

In semiconductors in-gap localized states lead to a photoconductivity effect
due to reduced relaxation rate of one of the PE carriers. Photoconductivity
was indeed observed in underdoped YBa$_{2}$Cu$_{3}$O$_{7-\delta }$ \cite
{drag}. With increase of the level of doping it is difficult to detect small
photoconductivity because of the increase of the dark conductivity. The
observation of long-lived component in photoinduced change of absorption is
more effective way to detect in-gap localized states because optical
conductivity in the range of charge transfer gap shows relatively small
changes with increase of doping.

\section{Theoretical model}

As photons from the pump laser pulse are absorbed, they excite electrons and
holes with the energy scale of the order of the photon energy. These
particles release their extra kinetic energy by electron-electron
scattering. When electrons reduce their energy to the value of the order of
phonon frequencies electron-phonon scattering becomes dominant. This energy
relaxation process is very rapid and the particles end up in QP states near
the Fermi energy within $\tau _{e}=$10$\sim $100 fs \cite{kdbm}. Subsequent
relaxation is slowed down by the presence of the gap and a relaxation
bottleneck is formed. From pump-probe photoinduced transmission experiments
in YBCO \cite{Han,kdbm} and BISCO \cite{Thomas}, the relaxation times of the
photoexcited QPs were found to be in the range $\tau _{QP}=$0.3$\sim $3 ps.
It is important that value of the gap in high-$T_{c}$ materials is of the
order of phonon frequencies, therefore electron-phonon scattering is
dominant only in recombination of electrons through the gap.

In addition to the picosecond transient, the signal on nanosecond timescale
has been consistently observed in HTSC \cite{Stevens,Thomas} and recently
also in the quasi 1D CDW insulator K$_{0.3}$MoO$_{3}$ \cite{bb}. After
photoexcitation the signal relaxes within 10 ps to some non-zero value, that
can be represented by a constant on the 100 ps timescale \cite{Thomas}. The
lifetime of the slow component, $\tau _{L}$, cannot be directly measured,
since it appears to be longer than the inter-pulse separation of $%
t_{r}\simeq 10-12$ ns. This results in a signal pile-up due to accumulation
of the response from many pulses (presence of photo-induced signal at
negative time-delay). From the experiments \cite{Stevens,bb} it appears that 
$\tau _{L}>10^{-7}$ s. This is long in comparison with the phonon relaxation
time and with the phonon escape time from the excitation volume into the
bulk or thin film substrate, which is typically 10$^{-10}$s, so we can
ignore phonon escape effects and discuss only intrinsic relaxation processes.

Since the relaxation time $\tau _{L}$ is long in comparison with $\tau _{QP}$
and phonon relaxation times, we assume that phonons and quasiparticles can
be described by equilibrium densities $N_{\omega }$ and $N$ respectively.
For the relaxation of the localized carriers we apply arguments similar to
those originally proposed by Rothwarf and Taylor \cite{Roth}. The rate
equation for the total density of localized excitations $N_{L}$ is then
given by: 
\begin{equation}
\frac{dN_{L}}{dt}=-RN_{L}^{2}-\tilde{\gamma}N_{L}+\gamma N+\beta N_{\omega }.
\end{equation}
The first term in Eq.(1) describes the recombination of two localized
excitations to a condensate with a recombination rate $R$. The second and
the third terms describe the exchange of an electron or a hole between the
localized and quasiparticle states with densities $N_{L}$ and $N$
respectively and with a rates $\tilde{\gamma}/\gamma \propto \exp {(-\Delta
E/k}_{B}{T)}$ where $\Delta E$ is the energy barrier between trapped
carriers and quasiparticles \cite{Ryvkin}. The last term describes the
spontaneous creation of localized excitations by phonons with a relaxation
rate $\beta $. When electron or hole is trapped additional contribution to
the energy barrier may appear due to a local lattice distortion (polaronic
effect). We assume that the energy barrier does not depend on whether
localized state is occupied or not.

Assuming the ansatz for $N_{L}=N_{L0}+n_{L},$ where $N_{L0}$ is equilibrium
density of localized particles and $n_{L}$ is the photoinduced density
created by the laser pulse and taking into account that $N_{\omega }$ and $N$
are given by their equilibrium values, we can rewrite Eq.(1): 
\begin{equation}
\frac{dn_{L}}{dt}=-Rn_{L}^{2}-(2RN_{L0}+\tilde{\gamma})n_{L}
\end{equation}
This equation is sufficiently general that it can be applied to different
systems with different types of ground states. The analytic solution of
Eq.(2) has the form: 
\begin{equation}
n_{L}(t)=\frac{n_{L}(0)}{(1+\frac{n_{L}(0)}{\tau })\exp (t/\tau )-\frac{%
n_{L}(0)R}{\tau }}.
\end{equation}
where $1/\tau =2N_{L0}R+\tilde{\gamma}$. Since we are interested in a steady
state solution for the case of excitation by repetitive pump pulse train, we
use the condition that total number of localized excitations that recombine
between two laser pulses should be equal to the number of localized
excitations created by each laser pulse. 
\begin{equation}
n_{L}(0)-n_{L}(t_{r})=\eta n_{QP}.
\end{equation}
Here $n_{QP}$ is the number of photoinduced quasiparticles created by each
laser pulse \cite{kdbm}, $\eta \propto \gamma \tau _{ph}$ is the probability
of trapping a photoinduced carrier into a localized state, $\tau
_{ph}\propto 1/\Delta (T)$ is quasiparticle relaxation time. Combining
Eqs.(3) and (4) one obtains the stationary solution for $n_{L}(0)$: 
\[
n_{L}(0)=\left[ \sqrt{1+\frac{\eta n_{QP}}{N_{L0}/2\tau t_{r}}}-1\right]
/2\tau R 
\]

\section{Results}

In the line with Ginzburg-Landau approach we can expand recombination rate $%
R $ in even powers of order parameter: 
\begin{equation}
R=\Gamma +\alpha \Delta (T)^{2}+..
\end{equation}
In the BCS-like case pairing occurs at T$_{c}$ and biparticle recombination
is impossible above T$_{c}$ ($\Gamma =0$). This result is consistent with
calculations of QP recombination rates\cite{Ovch}. In the case of preformed
pairing the first term in Eq.(6) is not 0 and biparticle recombination is
possible above T$_{c}$.

This simple observation makes temperature dependence of the amplitude of
slow component very different in these two cases \cite{slow}. In the case of
BCS-like pairing, the recombination rate is large at low temperature. It
leads to a small amplitude of the photoinduced signal. With increase of T,
the recombination rate decreases and relaxation via thermally activated QP
is dominant. As a result, a sharp maximum in the amplitude of the
photoinduced signal is observed near T$_{c}$ \cite{slow}. In the case of
preformed pairing, a similar increase of the amplitude takes place below T$%
_{c}$. Since $\Gamma $ is finite, the amplitude of the signal is a smooth
function above T$_{c}$ \cite{slow}.

To compare the theory (Eqs.(5,6)) with experiments we have plotted
experimental data for overdoped Ca$_{0.132}$Y$_{0.868}$Ba$_{2}$Cu$_{3}$O$%
_{6.928}$ (Fig.1a), for underdoped YBa$_{2}$Cu$_{3}$O$_{7-\delta }$ (Fig.1
b) and for CDW insulator K$_{0.3}$MoO$_{3}$ (Fig.2) samples. As clearly seen
from the fit theory provides qualitative and some time quantitative
description of the observed photoinduced changes of the absorption. In the
case of overdoped Ca$_{0.132}$Y$_{0.868}$Ba$_{2}$Cu$_{3}$O$_{6.928}$
(Fig.1a) a signal has been detected above T$_{c}$. This observation is
consistent with existence of pseudogap above T$_{c}$ \cite{Demsar} and
supports the statement that overdoped materials are spatially inhomogeneous.
In the case of underdoped YBa$_{2}$Cu$_{3}$O$_{7-\delta }$ (Fig.1 b) the
theory provides an accurate description of the data. The existence of a
pseudogap is clearly seen well above T$_{c}$.

The non-bolometric long-lived signal exists above T$_{c}$ in the case of CDW
insulator K$_{0.3}$MoO$_{3}$ (Fig.2) as well. This could be attributed to
the presence of pseudogap in the density of states due to a locally formed
gap. This is consistent with earlier measurements on K$_{0.3}$MoO$_{3}$ \cite
{gruener} that show a decrease of the density of states above T$_{c}$. In
all cases the parameters used in the fits are the same as used in the
analysis of QP relaxation \cite{kdbm,bb}. At low temperature there is a
deviation of the calculated curves from the experimental points. To explain
this effect we note that Eq.(6) is valid near T$_{c}$ where $\Delta $ is
small. Higher order terms in the expansion of $R$ can explain this deviation.

\section{Discussion}

In cuprate superconductors there is spectroscopic evidence suggesting that
there is a significant density of states in the gap possibly extending to
the Fermi level, which is often attributed to a $d$-wave gap symmetry.
However, by normal spectroscopies it is difficult to determine if the states
in the gap are QP states or, for example, localized states. Time-resolved
spectroscopy can answer this question rather effectively because of the
different time- and temperature-dependences of the QP and localized carrier
relaxations. It was argued that in the presence of impurity scattering QP
DOS in the $d$-wave state remains finite at zero energy \cite{Gorkov}.
Recently it was proposed \cite{Lee} that the quasiparticles in the
superconducting state may become localized for short coherence length $d$%
-wave superconductors. On the basis of available experimental data we cannot
make any definite conclusion about the \textit{origin} of intragap localized
states. However preliminary data on electron irradiated samples show only
weak dependence of the amplitude of the signal on the irradiation flux. It
is inconsistent with a simple $d$-wave interpretation of localized states
where an exponential dependence of the density of states on impurity
concentration is expected \cite{Gorkov}.

We can, however estimate the density of the intra-gap states from the
available data by assuming that the optical probe process is similar for
excited state absorption from localized states and for QPs. From typical
photoinduced reflection data for YBCO, we find that the amplitude of signal
associated with QP response is approximately equal to the amplitude of the
long-lived signal. It implies that also $n_{L}(0)\simeq n_{QP}$. From this
we can conclude that the density of intra-gap states is comparable with the
density of QP states. This observation has important implications for the
interpretation of frequency-domain spectroscopies, since it suggests that
the spectra should show a very significant intra-gap spectral density due to
localized states, irrespective of the gap symmetry.

In conclusion, we mention some of the most likely possibilities of the
origin of the localized states: (i) localized states associated with the
inhomogeneous ground state of the cuprates (stripes) \cite{stripes}, (ii)
intrinsic defect states, (iii) localized QP states in $d$-wave
superconductor \cite{Lee}. In K$_{0.3}$MoO$_{3}$, the nature intra-gap
excitations has been a subject of extensive study over the years and the
reader is referred to ref. \cite{gruener} for a review. However, the fact
that the signals in K$_{0.3}$MoO$_{3}$ and YBa$_{2}$Cu$_{3}$O$_{7-\delta }$
are very similar appears to rule out both spin excitations and vortex
states, leaving localized charges as the most plausible origin of the
intra-gap states.

\section{Figure Captions}

Figure 1: T-dependence of ns component amplitude $\mathcal{R}$ in a) Ca
overdoped YBCO and b) underdoped YBCO, together with the fits.

\medskip 

Figure 2: T-dependence of ns component amplitude $\mathcal{R}$ , taken on K$%
_{0.3}$MoO$_{3}$ (T$_{c}$ = 183 K) together with the fit using BCS-like gap.

\end{document}